\documentclass[aps,preprint,nofootinbib,eqsecnum]{revtex4}
\usepackage{amsmath}
\usepackage{amsfonts}
\usepackage{amssymb}
\usepackage{graphicx}
\usepackage{amscd}

\def\spur#1{\mathord{\not\mathrel{#1}}}

 \allowdisplaybreaks

\begin{document}

\begin{flushleft}
UCB-PTH-04/12 \\ 
LBNL-54938
\end{flushleft}

\title{Constraints and Superspin for SuperPoincar\'e Algebras in Diverse Dimensions}
\author{Andrea Pasqua and Bruno Zumino}
\affiliation{Lawrence Berkeley National Laboratory, 1 Cyclotron Rd., Berkeley, CA 94720,
USA}

\begin{abstract}
We generalize to arbitrary dimension the construction of a covariant and supersymmetric constraint for the
massless superPoincar\'e algebra, which was given for the eleven-dimensional case in a previous work. We also
contrast it with a similar construction appropriate to the massive case.
Finally we show that the constraint uniquely fixes the representation of the algebra.
\end{abstract}

\maketitle

\section{Constraints and Superspin}

We take as a starting point the superPoincar\'e algebra,\footnote{Notice that
the conventions we use in this paper differ slightly from those employed in \cite{BDPZ}.}
\begin{eqnarray}
&\left[ i J_{\rho \sigma}, P_\mu \right]&= \eta_{\mu \sigma} P_\rho -(\rho \leftrightarrow \sigma),\label{JP}\\
&\left[i J_{\rho\sigma}, J_{\mu\nu}\right]&= \left(\eta_{\mu\sigma} J_{\rho\nu}
- (\rho \leftrightarrow \sigma) \right) - (\mu \leftrightarrow \nu), \label{JJ} \\
&\left[ P_\mu, P_\nu \right]&=0, \quad \left[P_\mu, Q \right]=0 \\
&\left[ i J_{\rho\sigma}, Q \right]&=-\frac{1}{2}\Gamma_{\rho\sigma}Q,\label{JQ} \\ 
&\left\{Q,\bar{Q} \right\}&=-2 i \mathord{\not\mathrel{P}}.\label{QQbar}
\end{eqnarray}
Here, $\eta_{\mu\nu}=\textrm{diag}(-1, 1, \cdots , 1)$ and $\spur{P}=\Gamma_\mu P^\mu$, with $\Gamma_\mu$ satisfying the
Clifford algebra $\left\{\Gamma_\mu, \Gamma_\nu \right\}=2 \eta_{\mu\nu}$.
Also, $Q$ is a spinor of supercharge, $\bar{Q}=i Q^\dagger\Gamma^0$ and
$\Gamma_{\mu\nu}=\frac{1}{2} \left[\Gamma_\mu, \Gamma_\nu \right]$. All spinor indices are
suppressed; in particular $Q^\dagger=\left( Q^\ast \right)^T$ where $Q^\ast$ is the adjoint of $Q$ and $(\cdot)^T$ indicates
transposition with respect to the spinor indices.
Notice also that if $Q$ is a chiral spinor in $D$ spacetime dimensions, then the
right-hand side of (\ref{QQbar})
should contain a chiral projector $\frac{1\pm\Gamma}{2}$, and a convenient definition for $\Gamma$ is
$\Gamma=i^{D/2 -1} \Gamma_0 \Gamma_1 \cdots \Gamma_{D-1}$.
We define the supersymmetry variation of an operator $\mathcal{O}$ to be
\begin{equation}
\delta \mathcal{O}= \left\{Q,\mathcal{O} \right\} \; \textrm{or} \; \left[Q,\mathcal{O} \right],
\end{equation}
depending on whether $\mathcal{O}$ is fermionic or bosonic.

Next, we construct two antisymmetric three-tensors, namely\footnote{
In four dimensions, (\ref{W}) is the dual of the Pauli-Luba\'nski vector, so that $W$ should be thought of as its
generalization to higher dimensions.}
\begin{equation}
W_{\lambda\mu\nu}=P_{<\lambda}J_{\mu\nu >}=\frac{1}{3!}\sum_{Perm.} \pm P_{\lambda}J_{\mu\nu} \label{W},
\end{equation}
and
\begin{equation}
S_{\lambda\mu\nu}=\bar{Q}\Gamma_{\lambda\mu\nu}Q \label{S}.
\end{equation}
Here and in the following the angular brackets between indices indicate a sum over all permutations
of the indices, each taken with a sign and divided by the total number of permutations, as in (\ref{W}).
Furthermore, $\Gamma_{\lambda\mu\nu}=\Gamma_{<\lambda}\Gamma_{\mu}\Gamma_{\nu >}$.

Using the algebra (\ref{JP})-(\ref{QQbar}), it is easy to compute the supersymmetry variation of 
$W$. One finds
\begin{equation}
\delta W_{\lambda\mu\nu}=-\frac{i}{2} P_{<\lambda}\Gamma_{\mu\nu >} Q \label{deltaW}.
\end{equation}
As for $S$,
\begin{displaymath}
\delta S_{\lambda\mu\nu}=\delta \bar{Q}\Gamma_{\lambda\mu\nu}Q - \bar{Q}\Gamma_{\lambda\mu\nu} \delta Q.
\end{displaymath}
The first term in the expression above can be computed using (\ref{QQbar}).
The second term is different from zero only if $Q$ is a Majorana spinor. If it is
a Majorana spinor, $\delta Q$ can again be computed from (\ref{QQbar}) using the Majorana
condition\footnote{The Majorana condition takes the form $Q=BQ^\ast$ where $B$ is a matrix chosen in such a way that
$B^{-1}\Gamma_{\mu\nu}B=\Gamma_{\mu\nu}^\ast$. The Majorana condition can be imposed consistently in $D=2, 3, 4, 8, 9 \textrm{ mod } 8$
dimensions. In $D=2 \textrm{ mod } 8$ it can be imposed directly on a chiral spinor.}. The details
of the computation vary depending on the spacetime dimension and also, for even dimensions, on whether $Q$ is a
chiral or a Dirac spinor. But the final result is that, if the second term is nonzero, then it is exactly equal to the first term.
Hence, the supersymmetry variation of $S$ is twice as large when $Q$ is a Majorana spinor. In particular, we find
\begin{equation}
\delta S_{\lambda\mu\nu}= \left\{\begin{array}{ll}
				-2 i  \mathord{\not\mathrel{P}}\Gamma_{\lambda\mu\nu}Q, & \textrm{        if $Q$ is not a Majorana spinor},\\
				-4 i  \mathord{\not\mathrel{P}}\Gamma_{\lambda\mu\nu}Q, & \textrm{        if $Q$ is a Majorana spinor}.
				\end{array}
			   \right.
\label{deltaS}
\end{equation}

Now, if we are interested in massless representations of the superPoincar\'e algebra, it is convenient to rewrite the variation of $S$
as the sum of two terms, using the identity
\begin{equation}
\mathord{\not\mathrel{P}}\Gamma_{\lambda\mu\nu}= 6 P_{<\lambda}\Gamma_{\mu\nu >}- \Gamma_{\lambda\mu\nu} \mathord{\not\mathrel{P}}.
\label{massless}
\end{equation}
The first term has the same form as the variation of $W$ (\ref{deltaW}), and can be used to cancel it if we take an appropriate
linear combination of $W$ and $S$. The second term, instead, yields a variation proportional to $\mathord{\not\mathrel{P}}Q$, and 
$\mathord{\not\mathrel{P}}Q=0$ in a massless representation. We denote the relative
coefficient between $W$ and $S$ by $\kappa$ and their linear combination by $\Delta$,
\begin{equation}
\Delta_{\lambda\mu\nu}\equiv W_{\lambda\mu\nu}-\kappa \, S_{\lambda\mu\nu}. \label{Delta}
\end{equation}
It should be clear from the discussion above that the value of $\kappa$ depends only on whether $Q$ is or is not a Majorana spinor
and in particular
\begin{equation}
\kappa= \left\{\begin{array}{ll}
			\frac{1}{24} & \textrm{        if $Q$ is not a Majorana spinor},\\
			\frac{1}{48} & \textrm{        if $Q$ is a Majorana spinor}.
		\end{array}
	\right.
				\label{kappavalues}
\end{equation}
With the value of $\kappa$ as above, we find
\begin{equation}
\delta \Delta_{\lambda\mu\nu}=-\frac{1}{12}\Gamma_{\lambda\mu\nu} \mathord{\not\mathrel{P}}Q,
\end{equation}
so that it is possible to impose the constraints
\begin{equation}
P^2=0, \quad \mathord{\not\mathrel{P}}Q=0, \quad \Delta^{LMN}=0 \label{constraints}
\end{equation}
consistently with the full superPoincar\'e algebra\footnote{Actually, to impose consistently (\ref{constraints}),
one also needs $\delta \spur{P}Q \propto P^2$, which is true.}. The constraints (\ref{constraints}) were found in eleven-dimensional
spacetime in the course of the off-shell quantization of the superparticle \cite{BDPZ}, with the appropriate
value $\kappa=\frac{1}{48}$ for the relative coefficient ($Q$ is a Majorana spinor in eleven dimensions).
We will show in the next section that the constraint $\Delta$ actually fixes completely the representation and that in
eleven dimensions it fixes it to be the supergravity multiplet.

Two remarks. First, in the case of extended supersymmetry one can construct a tensor analogous to $\Delta_{\lambda\mu\nu}$ in a
straightforward way.
Namely, if there are $\mathcal{N}$ supercharges $Q_I$, then
\begin{equation}
\Delta_{\lambda\mu\nu}\equiv W_{\lambda\mu\nu}-\sum_{I=1}^\mathcal{N} \kappa_I \, \bar{Q}_I\Gamma_{\lambda\mu\nu}Q_I,  \label{extendedDelta}
\end{equation}
where each of the $\kappa_I$ is given by (\ref{kappavalues}). Then,
\begin{equation}
\delta_I \Delta_{\lambda\mu\nu} \equiv \left[ Q_I,\Delta_{\lambda\mu\nu}\right] = -\frac{1}{12}\Gamma_{\lambda\mu\nu}
\mathord{\not\mathrel{P}}Q_I,
\end{equation}
and again $\Delta$ can be set to zero consistently.
Secondly, in four spacetime dimensions the $\Delta$ tensor is only one of a continuous class of
supercovariant objects that can be constructed. Indeed, if one defines
\begin{equation}
\Delta_{\lambda\mu\nu}^{(\chi)} \equiv \Delta_{\lambda\mu\nu} - \frac{1}{3}\chi \, P^\alpha \epsilon_{\alpha\lambda\mu\nu},
	\label{Deltachi}
\end{equation} 
where $\chi$ is an arbitrary real number and $\epsilon$ is the completely antisymmetric tensor with $\epsilon_{0123}=+1$,
the supersymmetry variation of $\Delta^{(\chi)}$ is the same as that of $\Delta$. Hence $\Delta^{(\chi)}=0$ is also
a good constraint, compatible with the full superPoincar\'e algebra. We will elaborate on this point in the next section.

It is instructive to compare the $\Delta$-tensor to a similar construction which is useful for massive representations.
We could rewrite the variation (\ref{deltaS}) of $S$ using the identity
\begin{equation}
\mathord{\not\mathrel{P}}\Gamma_{\lambda\mu\nu}= 3 P_{<\lambda}\Gamma_{\mu\nu >}+
\frac{1}{2} \left[\mathord{\not\mathrel{P}}, \Gamma_{\lambda\mu\nu}\right],
\label{massive}
\end{equation}
instead of (\ref{massless}). Again the first term can be used to cancel the variation of $W$ if a suitable relative coefficient
between $W$ and $S$ is chosen. Let us emphasize that the coefficient needed differs from $\kappa$ by a factor of 2.
We call the linear combination $C_{\lambda\mu\nu}$ and the relative coefficient $\rho$. Hence,
\begin{equation}
C_{\lambda\mu\nu}\equiv W_{\lambda\mu\nu}-\rho \, S_{\lambda\mu\nu}, \label{C}
\end{equation}
with
\begin{equation}
\rho= \left\{\begin{array}{ll}
				\frac{1}{12} & \textrm{        if $Q$ is not a Majorana spinor},\\
				\frac{1}{24} & \textrm{        if $Q$ is a Majorana spinor}.
		\end{array}
	\right.
				\label{rhovalues}
\end{equation}
Then, 
\begin{equation}
\delta C_{\lambda\mu\nu}=-\frac{1}{12}\left[\Gamma_{\lambda\mu\nu}, \mathord{\not\mathrel{P}}\right]Q,
\end{equation}
and because of the identity $\left[P^\lambda \Gamma_{\lambda\mu\nu}, \mathord{\not\mathrel{P}}\right]=0$,
the antisymmetric tensor
\begin{equation}
C_{\mu\nu} \equiv P^\lambda C_{\lambda\mu\nu} \label{Cmunu}
\end{equation}
is invariant under supersymmetry transformations. Then, the scalar
\begin{equation}
C \equiv C_{\mu\nu} C^{\mu\nu} \label{superspin}
\end{equation}
is a Casimir of the full superPoincar\'e algebra and can be used to label its massive representations. For massless representations,
on the other hand, it is possible to show that $C$ vanishes identically\footnote{More precisely, one can show that in the frame
(\ref{lightconeP}), $C \propto A_i A^i$ where $A_i$ is given in (\ref{Ai}). As explained below, $A_i$ must vanish in a physically
sensible representation.}. In that case $\Delta$ is a more useful quantity to consider.
Notice that $C$ generalizes to arbitrary dimension a four-dimensional Casimir constructed in \cite{L, SS, S, PT}, where the
eigenvalues of that Casimir were termed ``superspin''.

\section{Nature of the Constraint $\Delta$}

To investigate how $\Delta=0$ constrains massless representations of the superPoincar\'e algebra, we choose a frame in which
$P=(E, E, 0, \cdots, 0)$ (light-cone frame). In $D$ spacetime dimensions, this choice breaks $SO(D-1, 1)$ down to
the ``little group'', $ISO(D-2)$,
namely the group of rotations and translations in $D-2$ dimensions. For convenience we introduce latin indices of two types,
$a, b, c =0, 1$ and $i, j, k=2, \cdots, D-1$, so that we can express the choice of frame with
\begin{equation} 
P^a=E, \; P^i=0.
\label{lightconeP}
\end{equation}
Then the components of $W$ are as follows,
\begin{eqnarray}
&W_{abc}&=W_{ijk} = 0, \label{Wabc}\\
&W_{abi}&=\epsilon_{ab}\frac{E}{3} \left(J_{i0}-J_{i1}\right) \equiv \epsilon_{ab}\frac{E}{3} A_i, \label{Ai} \\
&W_{aij}& = \pm \frac{E}{3} J_{ij},\label{Waij}
\end{eqnarray}
where $\epsilon_{ab}$ is the antisymmetric tensor in two dimensions with $\epsilon_{01}=+1$, and the upper sign in (\ref{Waij})
holds when $a=1$, the lower when $a=0$; similarly in (\ref{Saij}) and (\ref{Daij}) below.
Note that $A_i=\left(J_{i0}-J_{i1}\right)$ are precisely the generators of the translations of $ISO(D-2)$.

Before evaluating the components of $S$, we need to discuss how the frame choice (\ref{lightconeP}) affects the
supercharges $Q$, which, in a massless representation, are subject to the constraint $\spur{P}Q=0$. The answer
is that some components are projected out. Indeed
\begin{equation}
\pi_{+}Q=Q, \; \pi_{-}Q=0, \label{lightconeQ}
\end{equation}
where $\pi_{+}$ and $\pi_{-}$ are complementary projectors given by
\begin{equation}
\pi_{\pm}=\frac{1 \pm \Gamma^1 \Gamma^0}{2}.
\end{equation}
Using (\ref{lightconeQ}) and performing some algebra\footnote{As an example of the kind of derivations involved in computing
(\ref{Sabc})-(\ref{Saij}), we give a proof of \ref{Sabi}.
\begin{displaymath}
S_{abi}=\bar{Q}\Gamma_{abi}Q=\overline{(\pi_{+}Q)}\Gamma_{abi}\pi_{+}Q=
\bar{Q}\pi_{-}\Gamma_{abi}\pi_{+}Q=\bar{Q}\Gamma_{abi}\pi_{-}\pi_{+}Q=0.
\end{displaymath}.},
we see that the components of $S$ are
\begin{eqnarray}
&S_{abc}&=S_{ijk} = 0, \label{Sabc}\\
&S_{abi}&=0, \label{Sabi} \\
&S_{aij}&= \mp i Q^\dagger \Gamma_{ij}Q.\label{Saij}
\end{eqnarray}

Therefore, the components of $\Delta$ are
\begin{eqnarray}
&\Delta_{abc}&=\Delta_{ijk} = 0, \label{Dabc}\\
&\Delta_{abi}&=\epsilon_{ab}\frac{E}{3} A_i, \label{Dabi} \\
&\Delta_{aij}&=\pm \frac{E}{3}\left(J_{ij} + 3 i \, \kappa \frac{Q^\dagger \Gamma_{ij}Q}{E}\right),\label{Daij}
\end{eqnarray}
with $\kappa$ given by (\ref{kappavalues}).
We see that setting $\Delta=0$ is equivalent to imposing the pair of conditions
\begin{eqnarray}
			&A_i&=0,\label{Ai=0}\\
			&J_{ij}&= -3 i \, \kappa \, \frac{Q^\dagger \Gamma_{ij}Q}{E}.\label{Jeigenvalues}
\end{eqnarray}
The first condition requires that the translations of the little group be represented trivially. This is desireable on
physical grounds since a nontrivial representation would lead to unwanted continuous degrees of freedom, by a standard field theory
argument.
The second condition, on the other hand, restricts the eigenvalues of $J_{ij}$ to be those of the quadratic operator to the
right of (\ref{Jeigenvalues}). Those eigenvalues can be computed explicitly in any dimension. They are of course independent
of the values of $i$ and $j$, because the frame choice (\ref{lightconeP}) does not break the rotational invariance in the $i$
and $j$ indices. The eigenvalues are quantized as a result of the supersymmetry algebra (\ref{QQbar}). In this frame, the algebra
can also be written as\footnote{Again if the spinor $Q$ is chiral, one must add a chiral projector to the right-hand side of equation
(\ref{QQdagger}).}
\begin{equation}
		\left\{Q,Q^\dagger \right\}=4E \pi_{+}, \label{QQdagger} 
\end{equation}
from which it follows that the nonzero components of $Q$ are proportional to fermionic oscillators. How many oscillators exactly will
depend on the spacetime dimension and on what kind of spinor $Q$ is (for instance a chiral or a Majorana condition will each
reduce by half the number of independent oscillators). In the end, the right-hand side of (\ref{Jeigenvalues}) can be written
as a simple function of several fermionic number operators. Hence, the eigenvalues of $J$ and their multiplicities
can be easily computed and from that a representation can be inferred uniquely.

Shortly, we will give the form of that function of number operators for some interesting dimensions, including the eleven-dimensional
case of \cite{BDPZ}. But first two remarks are in order.
When extended supersymmetry is present (but with no central charges), all that we have done can be repeated with only minor changes.
The principal difference is that (\ref{Jeigenvalues}) is replaced by
\begin{equation}
		J_{ij}=-3 i \sum_{I=1}^\mathcal{N} \kappa_I \, \frac{Q_I^\dagger \Gamma_{ij}Q_I}{E},
\end{equation}
which in turn can be written as a function of $\mathcal{N}$ sets of number operators.
The second remark concerns the case of four spacetime dimensions, where a continuous class of constraints $\Delta^{(\chi)}$ exists,
as mentioned in the previous section. In four dimensions the little group is $ISO(2)$ and it consists of the helicity and of two
translations. With our choice of frame (\ref{lightconeP}), the generators are respectively $J_{23}$, $A_2$ and $A_3$.
Now, setting $\Delta^{(\chi)}=0$ adds a shift to the eigenvalues of $J_{23}$, namely to the helicities of the representation, while,
interestingly, the constraints $A_2=0$ and $A_3=0$ are unaffected,
\begin{equation}
		A_2=A_3=0,  \qquad J_{23}= -3 i \, \kappa \frac{Q^\dagger \Gamma_{23}Q}{E}+\chi.
\end{equation}
In summary, all possible representations with $A_2=A_3=0$ are recovered as $\chi$ varies. It should not come as a surprise that
$\chi$ appears to be a continuous variable, because our construction is purely algebraic whereas the quantization of the helicities
in four dimensions is a consequence of the topology of the little group\footnote{Namely, it follows from the fact that a
$4\pi$-rotation around the direction of the momentum can be continuously deformed into the identity.}.

As promised, we now give an expression for $J_{ij}$ in several interesting dimensions, being understood that $A_i$ is always zero.

Two and three spacetime dimensions are not interesting for our purposes since the little groups are either trivial or
consist only of a translation.
In four dimensions, when $Q$ is taken to be a chiral spinor or equivalently a Majorana spinor, there is only one independent fermionic
oscillator $a$ and the helicity operator is given by
\begin{equation}
	J_{23}[4D]=	\left\{ \begin{array}{ll}
				\chi \mp \frac{1}{2} a^\ast a, & \textrm{if $Q$ is left or right chiral respectively,}\\
				\chi +\frac{1}{4}- \frac{1}{2} a^\ast a, &  \textrm{if $Q$ is Majorana}.
				\end{array} 
			\right.
\label{4Drep}
\end{equation}
As claimed, (\ref{4Drep}) determines completely the representation of the superPoincar\'e algebra in terms of $\chi$.
It is given by the massless $\mathcal{N}=1$ supermultiplet in which the helicities are $\left(\chi \mp \frac{1}{2}, \chi \right)$
for a left or a right-handed chiral spinor, and $\left(\chi + \frac{1}{4}, \chi - \frac{1}{4}\right)$ for a Majorana spinor.
Now, in four dimensions it is a matter of convention whether $\mathcal{N}=1$ supersymmetry is implemented  with a left chiral,
a right chiral or a Majorana supercharge\footnote{Indeed, in four dimensions the complex conjugate of a left
chiral spinor is linearly related to a right chiral spinor and viceversa. Similarly a Majorana spinor can be constructed
in terms of a chiral spinor
and viceversa.}. Correspondingly, the way $\chi$ enters in the expression for the helicities depends
on that conventional choice but what matters is that in all cases each possible supermultiplet is recovered for an appropriate
value of $\chi$. For that value, $\Delta^{(\chi)}=0$ determines the representation in a fully supercovariant way.
It is intriguing that a supercovariant constraint for which $A_i \ne 0$ doesn't seem to exist. The generalization to the case
of extended supersymmetry is straightforward and again one recovers all possible representations for appropriate values of $\chi$.

In five dimensions, for $\mathcal{N}=1$, there are two independent fermionic oscillators $a_{1,2}$ and
\begin{equation}
		J_{ij}[5D]= \frac{1}{2} \left(a_2^\ast a_2 - a_1^\ast a_1 \right), \label{5Drep}
\end{equation}
which yields the supermultiplet with eigenvalues $\left(\frac{1}{2}, 0, 0, - \frac{1}{2} \right)$, namely one spin-$\frac{1}{2}$
and two spin zero particles, all three complex (indeed it is not possible to impose a Majorana condition on a spinor in five
dimensions). Under dimensional reduction, one obtains the $\mathcal{N}=2$ massless hypermultiplet in four dimensions.
It would be interesting to find supercovariant constraints that characterize the other representations of the superPoincar\'e algebra
in five or higher dimensions, as we were able to do in four dimensions.

Next, we consider ten dimensions. If we take $Q$ to be a Majorana-Weyl spinor we find that there are four independent
fermionic oscillators, $a_{1, \cdots, 4}$ and that
\begin{equation}
		J_{ij}[10D]= \frac{1}{2} \sum_{n=1}^4 a_n^\ast a_n - 1. \label{10Drep}
\end{equation}
Then, the eigenvalues for $J_{ij}$ are $(1,\frac{1}{2}, 0, -\frac{1}{2}, -1)$ with multiplicities $(1, 4, 6, 4, 1)$ and the
multiplet must be the gauge supermultiplet, which consists of an $SO(8)$ vector and an $SO(8)$ chiral spinor for a total
of $8$ fermionic and $8$ bosonic degrees of freedom. The generalization to extended supersymmetry is, once again, straightforward.
In particular for $\mathcal{N}=2$, we find the supergravity multiplets of type IIA or type IIB depending on the chiralities of the
two supercharges, as was to be expected.

Finally, in eleven dimensions with $Q$ Majorana, there are eight oscillators and
\begin{equation}
		J_{ij}[11D]= \frac{1}{2} \sum_{n=1}^8 a_n^\ast a_n - 2. \label{11Drep}
\end{equation}
The eigenvalues of $J_{ij}$ are $(\pm 2, \pm\frac{3}{2}, \pm 1, \pm \frac{1}{2}, 0)$ with multiplicities
$(1, 8, 28, 56, 70)$ for a total of $128$ fermionic and $128$ bosonic states, pointing inequivocably to the
eleven-dimensional supergravity multiplet.

\section{About the Massive Case}

For completeness, we present in this section a discussion of the tensor $C_{\mu\nu}$ of (\ref{Cmunu}). We proceed along the lines
of the discussion of $\Delta$.
What follows is a generalization to generic spacetime dimensions of similar arguments
that can be found in \cite{L, SS, S, PT} for the four-dimensional case.

To begin, we choose a frame in which $P=(m, 0, \cdots, 0)$ (rest frame). The little group is $SO(D-1)$ and it is generated by
$J_{ij}$ where $i, j = 1, \cdots, D-1$. In the rest frame, $\spur{P}=m\Gamma_0$ and the supersymmetry algebra (\ref{QQbar})
becomes\footnote{As usual, with a chiral projector on the right-hand side if $Q$ is chiral. The same for (\ref{aadagger}).}
$\left\{Q, Q^\dagger \right\}=2m$. If we rescale the supercharge $Q$, by defining $a \equiv \frac{Q}{\sqrt{2m}}$, then $a$ satisfies
\begin{eqnarray}
\left\{a, a^\dagger \right\}&=&1, \label{aadagger}\\
\left[ i J_{ij}, a \right]&=&-\frac{1}{2}\Gamma_{ij}a, \label{Ja}\\
\left[i J_{ij}, a^\dagger \right]&=& + \frac{1}{2} a^\dagger \Gamma_{ij}, \label{Jadagger}
\end{eqnarray}
where the latter two equations follow from (\ref{JQ}).
In the rest frame the components of $C_{\mu\nu}$ are
\begin{equation}
C_{0\mu}=0, \quad C_{ij}=-\frac{m^2}{3}\left[J_{ij} + 6  i \, \rho \, a^\dagger \Gamma_{ij} a \right], \label{Cijrestframe}
\end{equation}
with $\rho$ given by (\ref{rhovalues}).

We now define the tensors
\begin{equation}
T_{ij}= - 6i \, \rho \, a^\dagger \Gamma_{ij} a, \label{T}
\end{equation}
and
\begin{equation}
Y_{ij}=-\frac{3C_{ij}}{m^2}. \label{Y}
\end{equation}
The point of those definitions is that, as we will show presently, $T_{ij}$ and $Y_{ij}$ are angular momentum operators, in the sense that
they satisfy each the commutation relations of the generators of the little group $SO(D-1)$, exactly as $J_{ij}$ does.
Furthermore, $T$ and $Y$ commute with one another. Equation (\ref{Cijrestframe}) becomes
\begin{equation}
Y_{ij}=J_{ij} - T_{ij}, \textrm{ or equivalently } J_{ij}=Y_{ij} + T_{ij}, \label{composition}
\end{equation}
and therefore we can conclude that $J$ is the composition of two independent angular momentum operators, $T$ and $Y$.

To clarify the last statement, we should perhaps emphasize the following.
For the massive case,  giving a representation of $Y_{ij}$ amounts to fixing a particular representation of the
superPoincar\'e algebra. The reason is that the massive representations of the superPoincar\'e algebra are labelled by
$C=C_{\mu\nu}C^{\mu\nu}$, and, in the rest frame, $C$ is proportional to the quadratic Casimir associated with $Y_{ij}$,
$C \propto Y_{ij}Y^{ij}$.
Hence, in reading equation (\ref{composition}), we should keep in mind that the representation of $Y_{ij}$ is given.
Furthermore, we will argue below that the representation of $T_{ij}$ is fixed by its expression in terms of $a$. Which
representation exactly depends on the spacetime dimension and on what kind of spinor $Q$ is. At any rate, it will be a reducible
representation of $SO(D-1)$. 
Then equation (\ref{composition}) states precisely that $J_{ij}$ belongs to the tensor product of the given
representation of $Y_{ij}$ and the fixed representation of $T_{ij}$. The irreducible representations of $SO(D-1)$ contained in
that tensor product can be computed and together they form the supermultiplet associated with the given representation of
the superPoincar\'e algebra. We will illustrate this with an explicit example.

To find the commutation relations of $Y$ and $T$ one can proceed as follows.
The commutation relations of $T$ with $J$ can be computed using (\ref{Ja})-(\ref{Jadagger}). One finds that $T$
transforms as a tensor under $J$, namely
\begin{equation}
\left[i J_{ij}, T_{kl}\right]= \left(\eta_{kj} T_{il} - (i \leftrightarrow j) \right) - (k \leftrightarrow l).
\label{JT}
\end{equation}
The same is true for $Y=J-T$, since $J$ also transforms as a tensor, by (\ref{JJ}),
\begin{equation}
\left[i J_{ij}, Y_{kl}\right]= \left(\eta_{kj} Y_{il} - (i \leftrightarrow j) \right) - (k \leftrightarrow l).
\label{JY}
\end{equation}
Next, the commutator of $T$ with itself can be computed in terms of anticommutators of $a$ and $a^\ast$. More precisely
there will be terms containing the anticommutator of $a$ with $a^\ast$ and terms containing the anticommutator of $a$ with $a$ or
of $a^\ast$ with $a^\ast$. The first kind of terms, those involving anticommutators of $a$ with $a^\ast$, can be readily
computed using (\ref{aadagger}). The second kind of terms will vanish unless $Q$, and
therefore $a$, is a Majorana spinor. If $Q$ and $a$ are Majorana spinors, the anticommutators of $a$ with $a$ and
$a^\ast$ with $a^\ast$ can be computed using (\ref{aadagger})
together with the Majorana condition (see the note in the first section) and the result can be used to evaluate
the terms of the second kind. The details of the computation depend on
the spacetime dimensions and on whether $Q$ is chiral or not, but the end result is, once again, that the second kind of terms gives a
contribution exactly equal to that of the first kind of terms..
Therefore, we find
\begin{equation}
\left[iT_{ij}, T_{kl}\right]= \left\{\begin{array}{ll}
				(12\rho) \left(\eta_{kj} T_{il} - (i \leftrightarrow j) \right) - (k \leftrightarrow l) &
				\textrm{        if $Q$ is not a Majorana spinor},\\
				(24\rho) \left(\eta_{kj} T_{il} - (i \leftrightarrow j) \right) - (k \leftrightarrow l) &
				\textrm{        if $Q$ is a Majorana spinor},
				\end{array}
			   \right.
\end{equation}
and since $\rho$ is given by (\ref{rhovalues}) we find, as promised,
\begin{equation}
\left[i T_{ij}, T_{kl}\right]= \left(\eta_{kj} T_{il} - (i \leftrightarrow j) \right) - (k \leftrightarrow l).
\label{TT}
\end{equation}
Finally, it is easy to see that $Y$ and $T$ commute, by combining (\ref{JT}) and (\ref{TT}). Then the commutators of $Y$ with itself
follows from (\ref{JY}) and the fact that $Y$ and $T$ commute,
\begin{equation}
\left[ Y_{ij}, T_{kl}\right]= 0, \quad \left[i Y_{ij}, Y_{kl}\right]= \left(\eta_{kj} Y_{il} -
							(i \leftrightarrow j) \right) - (k \leftrightarrow l).
\label{YTYY}
\end{equation}
This proves the claim that $Y_{ij}$ and $T_{ij}$ are two commuting angular momentum operators.
 
To complete the discussion, we need only to present an argument that the representation of $T_{ij}$ is fixed.
The reasoning follows closely that given in the preceding section for $J_{ij}$ in the massless case. Indeed,
$T$ is quadratic in $a$ and $a^\ast$ and the components of $a$, $a^\ast$ are fermionic oscillators as a result of (\ref{aadagger}).
Therefore the eigenvalues of $T_{ij}$ can be obtained, together with their multiplicites, by rewriting $T_{ij}$
as a function of fermionic number operators. That in turn fixes the representation uniquely.
To illustrate this point, we consider one example, namely $D=4$ and $Q$ a chiral spinor.
The analysis is analogous to the five-dimensional massless case: the little group is $SO(3)$ and
there are two independent fermionic oscillators $a_1$, $a_2$. One finds
\begin{equation}
T_{12}=\frac{1}{2} (a_2^\ast a_2 - a_1^\ast a_1),
\end{equation}
with similar expressions for the other components of $T$. This fixes the eigenvalues of $T_{ij}$ to be
$(\frac{1}{2}, 0, 0, -\frac{1}{2})$. Hence, $T$ is determined to be a generator of the representation
$\mathbf{0}\oplus \mathbf{0} \oplus \mathbf{\frac{1}{2}}$. Then, if $Y_{ij}$ is chosen to be in a spin-$Y$ representation of $S0(3)$,
$J$ will be in $(\mathbf{0} \oplus \mathbf{0} \oplus \mathbf{\frac{1}{2}}) \otimes \mathbf{Y}= \mathbf{(Y-\frac{1}{2})}
\oplus \mathbf{Y} \oplus \mathbf{Y} \oplus \mathbf{(Y+\frac{1}{2})}$, namely in the $\mathcal{N}=1$ massive supermultiplet where the
highest spin is $Y+\frac{1}{2}$.

\section{Conclusion}
We introduced a covariant tensor  $\Delta$ which, in the case of a massless representation of the superPoincar\'e algebra,
is also supersymmetric. Imposing $\Delta=0$ is a supercovariant way to fix the representation completely, including the
generators of the translations in the little group. In particular, the translations are represented trivially,
as required on physical grounds.

For the case of nonvanishing mass, we have constructed an angular momentum operator $Y_{ij}$ and a Casimir $C$. The latter
generalizes to higher dimensions the superspin operator. We have also
shown how $Y_{ij}$ can be used to construct representations of the superPoincar\'e algebra.

The present investigation originated in work \cite{BDPZ} intended to define a superstar product appropriate for a
consistent formulation of supersymmetric string field theory. The results described here are interesting by themselves.
They show that (super)Lie algebras can admit constraints amenable to exact treatment and that statements about their
representations can be worked out.

After publication, it was pointed out to us that constraints and superspin
operators analogous to ours had been discussed earlier in the literature, in 
addition to the references we gave in our paper for the case of nonzero mass.
For four space-time dimensions, \cite{BK} gave also the case of vanishing mass.
Similar constraints, in general dimensions and for vanishing mass, were given
even earlier in a discussion of the dynamics of the superparticle and the
superstring from a superconformal point of view \cite{Sie}.

\section*{Aknowledgements}

We would like to thank Itzhak Bars, Mary K. Gaillard and David Olive for useful discussions. This work was supported in part
by the Director, Office of Science, Office of High Energy and Nuclear Physics, of the U.S. Department of Energy under
Contract DE-AC03-76SF00098, and in part by the NSF under grant 22386-13067.

\end{document}